\documentclass[twocolappendix,appendixfloats]{emulateapj}

\usepackage{apjfonts}
\usepackage{rotating}
\usepackage{amsmath}
\usepackage{color}
\usepackage{color}
\usepackage{CJK}

\newcommand{\pivec}{\mbox{\boldmath $\pi$}}
\newcommand{\muvec}{\mbox{\boldmath $\mu$}}
\newcommand{\te}{t_{\rm E}}
\newcommand{\thetae}{\theta_{\rm E}}
\newcommand{\pie}{\pi_{\rm E}}

\definecolor{darkbrown}{RGB}{139,69,19}

\shorttitle{OGLE-2017-BLG-0329}
\shortauthors{HAN ET AL.}

\begin{document}

\title{OGLE-2017-BLG-0329L: A Microlensing Binary Characterized with Dramatically Enhanced
Precision Using Data from Space-based Observations}

\author{
C.~Han\altaffilmark{101}, 
S.~Calchi Novati\altaffilmark{201,200},
A.~Udalski\altaffilmark{301,300}, 
C.-U.~Lee\altaffilmark{401,502,400},
A.~Gould\altaffilmark{401,402,403,400,200}, 
V.~Bozza\altaffilmark{202,800,500},\\
and\\
P.~Mr\'oz\altaffilmark{301}, 
P.~Pietrukowicz\altaffilmark{301}, 
J.~Skowron\altaffilmark{301}, 
M.~K.~Szyma\'nski\altaffilmark{301}, 
R.~Poleski\altaffilmark{301,402}, 
I.~Soszy\'nski\altaffilmark{301},
S.~Koz{\l}owski\altaffilmark{301}, 
K.~Ulaczyk\altaffilmark{301}, 
M.~Pawlak\altaffilmark{301}, 
K.~Rybicki\altaffilmark{301},
P.~Iwanek\altaffilmark{301},\\
(The OGLE Collaboration) \\   
M.~D.~Albrow\altaffilmark{501}, 
S.-J.~Chung\altaffilmark{401,502}, 
K.-H.~Hwang\altaffilmark{401}, 
Y.~K.~Jung\altaffilmark{401,503},
Y.-H.~Ryu\altaffilmark{401},
I.-G.~Shin\altaffilmark{503},
Y.~Shvartzvald\altaffilmark{204,600},
J. C.~Yee\altaffilmark{503,200},
W.~Zang\altaffilmark{404,405},
W.~Zhu\altaffilmark{406},
S.-M.~Cha\altaffilmark{401,504}, 
D.-J.~Kim\altaffilmark{401}, 
H.-W.~Kim\altaffilmark{401}, 
S.-L.~Kim\altaffilmark{401,502}, 
D.-J.~Lee\altaffilmark{401},
Y.~Lee\altaffilmark{401,504}, 
B.-G.~Park\altaffilmark{401,502}, 
R.~W.~Pogge\altaffilmark{402},
W.-T.~Kim\altaffilmark{505} \\ 
(The KMTNet Collaboration),\\
C.~Beichman\altaffilmark{203}, 
G.~Bryden\altaffilmark{204}, 
S.~Carey\altaffilmark{205}, 
B.~S.~Gaudi\altaffilmark{402}, 
C.~B.~Henderson\altaffilmark{204},\\
(The $Spitzer$ Team)\\
M.~Dominik\altaffilmark{801},
C.~Helling\altaffilmark{801},
M.~Hundertmark\altaffilmark{802},
U.~G.~J{\o}rgensen\altaffilmark{803},
P.~Longa-Pe{\~n}a\altaffilmark{804},
S.~Lowry\altaffilmark{805},
S.~Sajadian\altaffilmark{806},
M.~J.~Burgdorf\altaffilmark{807},
J.~Campbell-White\altaffilmark{805},
S.~Ciceri\altaffilmark{808},
D.~F.~Evans\altaffilmark{809},
L.~K.~Haikala\altaffilmark{810},
T.~C.~Hinse\altaffilmark{811},
S.~Rahvar\altaffilmark{812},
M.~Rabus\altaffilmark{813,814},
C.~Snodgrass\altaffilmark{815} \\
(The MiNDSTEp Collaboration)
}

\email{cheongho@astroph.chungbuk.ac.kr}

\altaffiltext{101}{Department of Physics, Chungbuk National University, Cheongju 28644, Republic of Korea} 
\altaffiltext{201}{IPAC, Mail Code 100-22, California Institute of Technology, 1200 E. California Boulevard, Pasadena, CA 91125, USA} 
\altaffiltext{202}{Dipartimento di Fisica "E. R. Caianiello", Universit\'a Salerno, Via Giovanni Paolo II, I-84084 Fisciano (SA), Italy} 
\altaffiltext{301}{Warsaw University Observatory, Al. Ujazdowskie 4, 00-478 Warszawa, Poland} 
\altaffiltext{401}{Korea Astronomy and Space Science Institute, Daejon 34055, Republic of Korea} 
\altaffiltext{402}{Department of Astronomy, Ohio State University, 140 W. 18th Ave., Columbus, OH 43210, USA} 
\altaffiltext{403}{Max Planck Institute for Astronomy, K\"onigstuhl 17, D-69117 Heidelberg, Germany} 
\altaffiltext{404}{Physics Department and Tsinghua Centre for Astrophysics, Tsinghua University, Beijing 100084, China}
\altaffiltext{405}{Department of Physics, Zhejiang University, Hangzhou, 310058, China}
\altaffiltext{406}{Canadian Institute for Theoretical Astrophysics, University of Toronto, 60 St George Street, Toronto, ON M5S 3H8, Canada}
\altaffiltext{501}{University of Canterbury, Department of Physics and Astronomy, Private Bag 4800, Christchurch 8020, New Zealand} 
\altaffiltext{502}{Korea University of Science and Technology, 217 Gajeong-ro, Yuseong-gu, Daejeon, 34113, Republic of Korea} 
\altaffiltext{503}{Harvard-Smithsonian Center for Astrophysics, 60 Garden St., Cambridge, MA 02138, USA} 
\altaffiltext{504}{School of Space Research, Kyung Hee University, Yongin, Kyeonggi 17104, Korea} 
\altaffiltext{505}{Department of Physics \& Astronomy, Seoul National University, Seoul 08826, Republic of Korea} 
\altaffiltext{203}{NASA Exoplanet Science Institute, California Institute of Technology, Pasadena, CA 91125, USA} 
\altaffiltext{204}{Jet Propulsion Laboratory, California Institute of Technology, 4800 Oak Grove Drive, Pasadena, CA 91109, USA} 
\altaffiltext{205}{$Spitzer$ Science Center, MS 220-6, California Institute of Technology, Pasadena, CA, USA} 
\altaffiltext{800}{Istituto Nazionale di Fisica Nucleare, Sezione di Napoli, Via Cintia, 80126 Napoli, Italy} 
\altaffiltext{801}{Centre for Exoplanet Science, SUPA, School of Physics \& Astronomy, University of St Andrews, North Haugh, St Andrews KY16 9SS, UK}
\altaffiltext{802}{Astronomisches Rechen-Institut, Zentrum f{\"u}r Astronomie der Universit{\"a}t Heidelberg (ZAH), 69120 Heidelberg, Germany}
\altaffiltext{803}{Niels Bohr Institute \& Centre for Star and Planet Formation, University of Copenhagen, {\O}ster Voldgade 5, 1350 Copenhagen, Denmark}
\altaffiltext{804}{Unidad de Astronom{\'{\i}}a, Universidad de Antofagasta, Av.\ Angamos 601, Antofagasta, Chile}
\altaffiltext{805}{Centre for Astrophysics \&\ Planetary Science, The University of Kent, Canterbury CT2 7NH, UK}
\altaffiltext{806}{Department of Physics, Isfahan University of Technology, Isfahan 84156-83111, Iran}
\altaffiltext{807}{Universit{\"a}t Hamburg, Faculty of Mathematics, Informatics and Natural Sciences, Department of Earth Sciences, Meteorological Institute, Bundesstra\ss{}e 55, 20146 Hamburg, Germany}
\altaffiltext{808}{Department of Astronomy, Stockholm University, Alba Nova University Center, 106 91, Stockholm, Sweden}
\altaffiltext{809}{Astrophysics Group, Keele University, Staffordshire, ST5 5BG, UK}
\altaffiltext{810}{Instituto de Astronomia y Ciencias Planetarias de Atacama,  Universidad de Atacama, Copayapu 485,  Copiapo, Chile}
\altaffiltext{811}{Korea Astronomy \& Space Science Institute, 776 Daedukdae-ro, Yuseong-gu, 305-348 Daejeon, Republic of Korea}
\altaffiltext{812}{Department of Physics, Sharif University of Technology, PO Box 11155-9161 Tehran, Iran}
\altaffiltext{813}{Instituto de Astrof\'isica, Pontificia Universidad Cat\'olica de Chile, Av. Vicu\~na Mackenna 4860, 7820436 Macul, Santiago, Chile}
\altaffiltext{814}{Max Planck Institute for Astronomy, K{\"o}nigstuhl 17, 69117 Heidelberg, Germany}
\altaffiltext{815}{School of Physical Sciences, Faculty of Science, Technology, Engineering and Mathematics, The Open University, Walton Hall, Milton Keynes, MK7 6AA, UK} 
\altaffiltext{200}{$Spitzer$ Team.}
\altaffiltext{300}{OGLE Collaboration.}
\altaffiltext{400}{KMTNet Group.}
\altaffiltext{500}{MiNDSTEp Collaboration.}
\altaffiltext{600}{NASA Postdoctoral Program Fellow.}

\begin{abstract}
Mass measurements of gravitational microlenses require one to determine the microlens parallax $\pie$, 
but precise $\pie$ measurement, in many cases, is hampered due to the subtlety of the microlens-parallax 
signal combined with the difficulty of distinguishing the signal from those induced by other higher-order 
effects.  In this work, we present the analysis of the binary-lens event OGLE-2017-BLG-0329, for which 
$\pie$ is measured with a dramatically improved precision using additional data from space-based $Spitzer$ 
observations. We find that while the parallax model based on the ground-based data cannot be distinguished 
from a zero-$\pie$ model at 2$\sigma$ level, the addition of the $Spitzer$ data enables us to identify 2 
classes of solutions, each composed of a pair of solutions according to the well-known ecliptic degeneracy. 
It is found that the space-based data reduce the measurement uncertainties of the north and east components 
of the microlens-parallax vector $\pivec_{\rm E}$ by factors $\sim 18$ and $\sim 4$, respectively. With the 
measured microlens parallax combined with the angular Einstein radius measured from the resolved caustic 
crossings, we find that the lens is composed of a binary with components masses of either 
$(M_1,M_2)\sim (1.1,0.8)\ M_\odot$ or $\sim (0.4,0.3)\ M_\odot$ according to the two solution classes.  
The first solution is significantly favored but the second cannot be securely ruled out based on the 
microlensing data alone.  However, the degeneracy can be resolved from adaptive optics observations 
taken $\sim 10$ years after the event.
\end{abstract}

\keywords{gravitational lensing: micro  -- binaries: general}

\section{Introduction}\label{sec:one}
Microlensing phenomena occur by the gravitational field of lensing objects regardless of their luminosity.
Due to this nature, microlensing can, in principle, provide an important tool to determine the mass
spectrum of Galactic objects based on samples that are unbiased by luminosity \citep{Han1995}.

Construction of the mass spectrum requires one to determine the masses of individual lenses. For most
microlensing events, the only observable related to the physical parameters of the lens is the Einstein
timescale.  However, the Einstein timescale is related to not only the lens mass but also the relative
lens-source parallax, $\pi_{\rm rel}$, and the proper motion, $\mu_{\rm rel}$, by
\begin{equation}
t_{\rm E}= {\thetae\over \mu_{\rm rel}},\qquad  \thetae=(\kappa M \pi_{\rm rel})^{1/2},
\end{equation}
where $\thetae$ is the angular Einstein radius, 
$\kappa=4G/(c^2 {\rm au})\sim 8.14\ {\rm mas}/M_\odot$, 
$\pi_{\rm rel}={\rm au}(D_{\rm L}^{-1}-D_{\rm S}^{-1})$, 
and $D_{\rm L}$ and $D_{\rm S}$ denote the distances to the lens and source, respectively.  As a result, 
the lens mass cannot be uniquely determined from the event timescale alone. For the unique determination 
of the lens mass, one needs to measure two additional quantities: the angular Einstein radius $\thetae$ 
and the microlens-parallax $\pie$ with which the mass and distance to the lens are determined by
\citep{Gould2000}
\begin{equation}
M={\thetae\over \kappa \pie};\qquad
D_{\rm L}={{\rm au}\over \pie\thetae+\pi_{\rm S}}
\end{equation}
where $\pi_{\rm S}={\rm au}/D_{\rm S}$.

The angular Einstein radius can be measured from deviations in lensing lightcurves affected by finite-source
effects. Finite-source effects occur when a source star is located in the region within which the gradient of
lensing magnifications is significant and thus different parts of the source are differentially magnified. For a
lensing event produced by a single mass, this corresponds to the very tiny region around the lens, and thus
finite-source effects can be effectively detected only for a very small fraction of events for which the lens 
transits the surface of the source \citep{Witt1994, Nemiroff1994, Gould1994}.  For events produced by binary 
lenses, on the other hand, the chance to detect finite-source effects is relatively high because the lens 
systems form extended caustics around which the magnification gradient is high. Analysis of deviations 
affected by finite-source effects yields the normalized source radius $\rho$, which is defined as the ratio 
of the angular source radius  $\theta_*$ to the angular Einstein radius. By estimating $\theta_*$ from 
external information of the source color, the angular Einstein radius is determined by $\thetae=\theta_*/\rho$.

The microlens-parallax can be measured from deviations in lensing lightcurves caused by the positional
change of an observer. In the single frame of Earth, such deviations occur due to the acceleration of
Earth induced by the orbital motion: ``annual microlens parallax'' \citep{Gould1992}. However,
precise $\pie$ measurement from the deviations induced by the annual microlens-parallax effect is difficult
because the positional change of an observer during $\sim (O)10$-day durations of typical lensing events is, 
in most cases, very minor. As a result, $\pie$ measurements have been confined to 
a small fraction of all events, preferentially events with long timescales and/or events
caused by relatively nearby lenses.
For binary-lens events, $\pie$ measurement 
becomes further complicated because the orbital motion of the binary lens produces deviations in lensing 
light curves similar to those induced by microlens-parallax effects \citep{Batista2011, Skowron2011, Han2016}.

Microlens parallaxes of lensing events can also be measured if events are simultaneously observed using
ground-based telescopes and space-based satellite in a heliocentric orbit: ``space-based microlens parallax''
\citep{Refsdal1966, Gould1994}. For typical lensing events with physical Einstein radii of order au, 
the separation between Earth and a satellite can comprise
a significant fraction of the Einstein radius. Then, the lensing 
lightcurves observed from the ground and from the satellite appear to be different due to the difference 
in the relative lens-source positions, and the comparison of the two lightcurves leads to the determination 
of $\pie$.

In this work, we present the analysis of the binary microlensing event OGLE-2017-BLG-0329 that was
observed both from the ground and in space using the $Spitzer$ telescope. We show that while the parallax
model based on the ground-based data cannot be distinguished from a zero-$\pie$ model, the addition of the
$Spitzer$ data leads to the firm identification of two classes of microlens-parallax solutions

\begin{figure*}
\epsscale{0.85}
\plotone{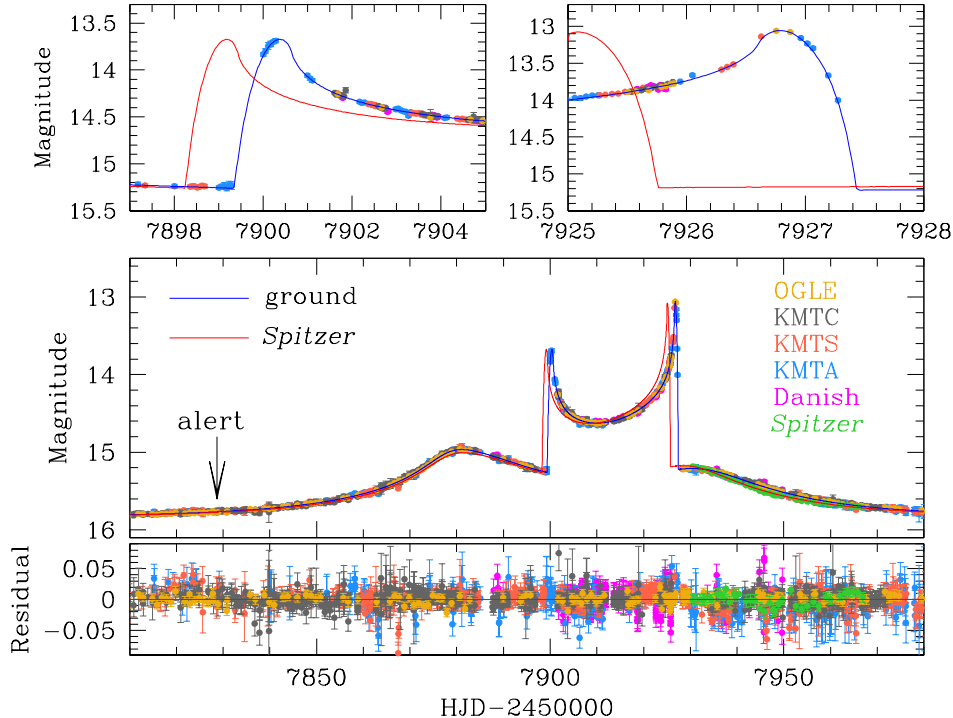}
\caption{
Light curve of OGLE-2017-BLG-0329. The blue and red curves superposed on the data points represent the 
best-fit model light curves for the ground and space-based data, respectively. The arrow designates the 
time when the event was alerted. The upper panels show the enlarged views of the caustic entrance (left 
panel) and exit (right panel) parts of the light curve.
}
\label{fig:one}
\end{figure*}

\section{Observations and Data}\label{sec:two}

The microlensing event OGLE-2017-BLG-0329 occurred on a star located toward the Galactic bulge. 
The equatorial coordinates of the event are $({\rm RA},{\rm DEC})_{\rm J2000}=$(17:53:43.20, -32:55:27.4), 
which correspond to the Galactic coordinates $(l,b)=(-2.53^\circ,-3.54^\circ)$. The baseline 
magnitude of the event before lensing magnification was $I_{\rm base}\sim 15.84$.

Figure~\ref{fig:one} shows the light curve of the event. The light curve is characterized by 3
peaks. The first smooth peak occurred at ${\rm HJD}'={\rm HJD}-2450000 \sim 7882$ and the other 
two sharp peaks occurred at ${\rm HJD}' \sim 7900$ and 7927.  The smooth and sharp peaks are 
typical features that occur when a source approaches the cusp and passes over the fold of a 
binary-lens caustic, respectively.  The event was already in progress before the 2017 microlensing 
season and lasted for more than 100 days.

The lensing event was observed from the ground by two microlensing surveys of the Optical Gravitational
Lensing Experiment \citep[OGLE:][]{Udalski2015a} and the Korea Microlensing Telescope Network
\citep[KMTNet:][]{Kim2016}. OGLE observations of the event were conducted using the 1.3m telescope
located at the Las Campanas Observatory in Chile.  The OGLE survey first identified the event from 
its Early Warning System on 2017 March 14 (${\rm HJD}'=7828.4$).  KMTNet observations were carried 
out using 3 globally distributed 1.6m telescopes located at the Cerro Tololo Inter-American Observatory 
in Chile (KMTC), the South African Astronomical Observatory in South Africa (KMTS), and the Siding 
Spring Observatory in Australia (KMTA).  
The event was identified by KMTNet as BLG22K0103.001613.
Observations by both surveys were conducted mainly in $I$ 
band and some $V$-band images were obtained for the color measurement of the source star.  
The event was located in the OGLE BLG502 and KMTNet BLG22 fields, which were observed with 
cadences of 0.17/hr and 1/hr by the OGLE and KMTNet surveys, respectively. 
With the high cadence of the surveys, both the caustic entrance and exit were resolved. 
See the upper panels of Figure~\ref{fig:one}.
Besides the survey experiments, the event was additionally observed from follow-up 
experiment conducted by the MiNDSTEp Collaboration during the period 
$7887.9 < {\rm HJD}' < 7954.7$ using the 1.5m Danish Telescope at La Silla Observatory in Chile.
Photometry 
of the data were conducted using the pipelines developed by the individual groups 
based on the difference imaging analysis method \citep{Alard1998}.  Since the data 
sets were taken using different instruments and reduced based on different softwares, we normalize 
the error bars of the individual data sets using the method described in \citet{Yee2012}.

The event was also observed in space.  
At the time that it was originally evaluated for {\it Spitzer} observations
(2017 May 1; ${\rm HJD}' = 7874$), it was believed to be a point-lens event,
and hence the decision was made in accordance with the protocols of
Yee et al. (2015), which are designed to obtain an unbiased sample
of events to probe the Galactic distribution of planets.
The {\it Spitzer} team specified that the event should be observed
provided that it reached $I<15.65$ at ${\rm HJD}'=7924$, i.e., the time
of the first upload. Since this requirement was met, these observations
were initiated, and were ultimately conducted during
the period 7930.5 -- 7966.9 ($\sim 36.4$ days), 
with both dates set essentially by the spacecraft's Sun-angle restrictions.
$Spitzer$ images were taken in the 3.6 $\mu$m channel of the IRAC camera, and the data were reduced 
with a specially developed version of point response function photometry \citep{Calchi2015b}.

\section{Analysis}\label{label:three}

OGLE-2017-BLG-0329 is of scientific importance because it may be possible to measure the microlens
parallax not only from the ground-based data (annual microlens parallax) but also from the combined 
ground+space data (space-based microlens parallax).  For this event, the chance to measure the 
annual microlens parallax is high due to its long timescale. Since the event was 
additionally observed by the $Spitzer$ telescope, the microlens parallax can also be measured from 
the combined ground+space data. Therefore, the event provides a test bed in which one can check the 
consistency of the $\pie$ values and compare the precision of $\pie$ measurements by the individual 
methods.  We note that there exist four cases for which ground-based $\pie$ measurements have been 
confirmed by space-based data: OGLE-2014-BLG-0124 \citep{Udalski2015b}, OGLE-2015-BLG-0196 \citep{Han2017}, 
OGLE-2016-BLG-0168 \citep{Shin2017}, and MOA-2015-BLG-020 \citep{Wang2017}. 


\subsection{Ground-based Data}\label{sec:three-one}

We first conduct analysis of the event based on the data obtained from ground-based observations. We start
modeling the light curve under the approximation that the relative lens-source motion is rectilinear (``standard
model''). For this modeling, one needs 7 principal lensing parameters. These parameters include the time of
the closest source approach to a reference position of the binary lens, $t_0$, the lens-source separation at 
that time, $u_0$ (impact parameter), the event timescale, $t_{\rm E}$, the projected separation $s$ (normalized 
to $\thetae$), and the mass ratio $q$ between the binary-lens components, the angle between the source trajectory 
and the binary-lens axis, $\alpha$ (source trajectory angle), and the normalized source radius $\rho$. 
We choose the barycenter of the binary lens as the reference position of the lens.

Since both the caustic crossings of the light curve were resolved, we consider finite-source effects.
Finite-source magnifications are computed using the ray-shooting method
\citep{Schneider1986, Kayser1986, Wambsganss1997}
In computing lensing magnifications,
we consider the surface-brightness variation of the source star caused by limb darkening. For this, we
model the surface brightness profile of the source star as $S \propto 1-\Gamma(1-3\cos \phi/2)$, where 
$\Gamma$ is the linear limb-darkening coefficient and $\phi$ is the angle between the line of sight toward 
the center of the source star and the normal to the surface of the source star. Based on the spectral type 
of the source star (see Section~\ref{sec:four-one}), we adopt $\Gamma_I=0.53$.

To find the solution of the lensing parameters, we first conduct a grid search for the parameters 
$\log s$ and $\log q$, while the other parameters $(t_0, u_0, t_{\rm E}, \rho, \alpha)$ at each point 
on the ($\log s,\log q$) plane are searched for by minimizing $\chi^2$ using the Markov Chain Monte 
Carlo (MCMC) method.  This first-round search yields local minima in the ($\log s,\log q$) plane.  
For each local minimum, we then refine the solution by allowing all parameters to vary. We identify 
a global minimum by comparing $\chi^2$ values of the individual local solutions. From this modeling, 
we find a unique solution of the event. According to this solution, the event was produced by a binary 
with a mass ratio between the components of $q\sim 0.7$ and a projected separation of $s\sim 1.4$. 
Due to the similar masses of the binary components and the proximity of the separation to unity, the 
caustic forms a single big closed curve (resonant caustic).

\begin{deluxetable}{lcc}
\tablecaption{Comparison of models (Ground-based data) \label{table:one}}
\tablewidth{0pt}
\tabletypesize{\small}
\tablehead{
\multicolumn{1}{c}{Model}  &
\multicolumn{2}{c}{$\chi^2$}  \\
\multicolumn{1}{c}{} &
\multicolumn{1}{c}{$u_0>0$ solution}  &
\multicolumn{1}{c}{$u_0<0$ solution}  
}
\startdata                                              
Standard               &  2392.1      &  -        \\   
Orbit                  &  2338.2      &  -        \\   
Parallax               &  2356.8      &  2363.1   \\   
Orbit+Parallax         &  2336.1      &  2330.8         
\enddata                                              
\end{deluxetable}

Since the event can be subject to higher-order effects due to its long timescale, we conduct 
additional modeling considering two such effects.  In the ``parallax model'' and ``lens-orbital 
model'', we separately consider the microlens-parallax and lens-orbital effects, respectively. We also 
conduct modeling by simultaneously considering both higher-order effects (``orbit+parallax model'').  
Consideration of the microlens-parallax effects requires to include 2 additional parameters of 
$\pi_{{\rm E},N}$ and $\pi_{{\rm E},E}$ , which represent the north and east components of the microlens 
parallax vector $\pivec_{\rm E}$, projected on the sky in the north and east equatorial coordinates, 
respectively. Under the approximation that the positional change of the lens is small, the lens-orbital 
effects are described by two parameters $ds/dt$ and $d\alpha/dt$, which represent the change rates of 
the binary separation and the source trajectory angle, respectively.  For parallax solutions, it is 
known that there can exist a pair of degenerate solutions with $u_0>0$ and $u_0<0$ due to the mirror 
symmetry of the lens system geometry \citep{Smith2003, Skowron2011}. We check this so-called ``ecliptic 
degeneracy'' whenever we consider microlens-parallax effects in modeling. The lensing parameters of the 
two solutions resulting from the ecliptic degeneracy are approximately in the relations of 
$(u_0, \alpha, \pi_{{\rm E},N}, d\alpha/dt) \leftrightarrow -(u_0, \alpha, \pi_{{\rm E},N}, d\alpha/dt)$.

In Table~\ref{table:one}, we present the goodness of the fits expressed in terms of $\chi^2$ values for 
the individual tested models. From the comparison of $\chi^2$ values, it is found that the model fit 
improves with the consideration of the higher-order effects. The improvement by the microlens-parallax 
and lens-orbital effects are $\chi^2=35.3$ and 53.9, respectively.  When both higher-order effects are 
simultaneously considered, on the other hand, it is found that the fit improvement is merely $\chi^2=7.4$ 
with respect to the orbital model. From the fact that (1) the fit improvement by the lens-orbital effect 
is bigger than the improvement by the microlens-parallax effect and (2) the further improvement from the 
orbital model by additionally considering the microlens-parallax effect is small, we judge that the 
dominant higher-order effect is the lens-orbital effect and the microlens-parallax effect is relatively 
small.

\begin{figure}
\epsscale{0.80}
\plotone{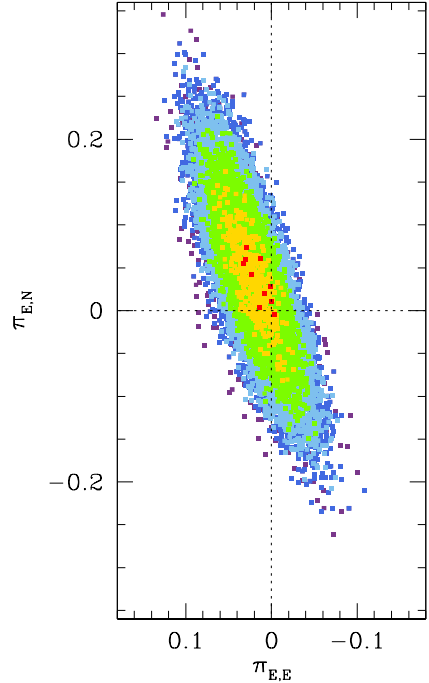}
\caption{
Distribution of $\Delta\chi^2$ in the ($\pi_{{\rm E},N},\pi_{{\rm E},E}$) plane obtained
from the analysis based on the ground-based data. Color coding indicates points in 
the MCMC chain within $1\sigma$ (red), $2\sigma$ (yellow), $3\sigma$ (green), $4\sigma$ 
(cyan), and $5\sigma$ (blue).
}
\label{fig:two}
\end{figure}

The weakness of the microlens-parallax effect can also be seen in Figure~\ref{fig:two}, where we present 
the $\Delta\chi^2$ distribution in the ($\pi_{{\rm E},N},\pi_{{\rm E},E}$) plane obtained from the modeling 
considering both microlens-lens and lens-orbital effects.  It shows that the model is consistent with the 
zero-$\pie$ model by $\Delta\chi^2\lesssim 4$.  For the validation of the weak microlens-parallax 
interpretation, the lens parameters resulting from the orbit+parallax model should be physically 
permitted.  For this, we estimate the ranges of the lens mass ($M=M_1+M_2$) and the projected 
kinetic-to-potential energy ratio, which is computed by 
\begin{equation}
\left({{\rm KE} \over {\rm PE}} \right)_\perp = 
{(a_\perp/{\rm au})^3 \over 8\pi (M/M_\odot)}
\left[ \left( {1\over s} {ds\over dt}{\rm yr} \right)^2 + \left( { d\alpha\over dt} {\rm yr}\right)^2 \right] .
\end{equation}
We describe the procedure to measure the angular Einstein radius $\thetae$, which is needed to determine 
$M$ and $({\rm KE/PE})_\perp$, in Section~\ref{sec:four-one}.  We find that the ranges of the lens mass 
and the energy 
ratio are $0.9 \leq  M/M_\odot \leq 4.6 $ and $0.04 \leq ({\rm KE/PE})_\perp \leq 0.1$, respectively.  
The estimated lens mass roughly corresponds to those of binaries composed of stars.  The kinetic-to-potential 
energy ratio also meets the condition $({\rm KE}/{\rm PE})_\perp < {\rm KE}/{\rm PE}< 1$, that is required 
for the binary lens to be a gravitationally bound system.  Therefore, the solution based on the ground-based 
data is physically permitted, although the range of the estimated lens mass is very wide due to the large 
uncertainty of the measured $\pie$.

\begin{figure}
\epsscale{0.80}
\plotone{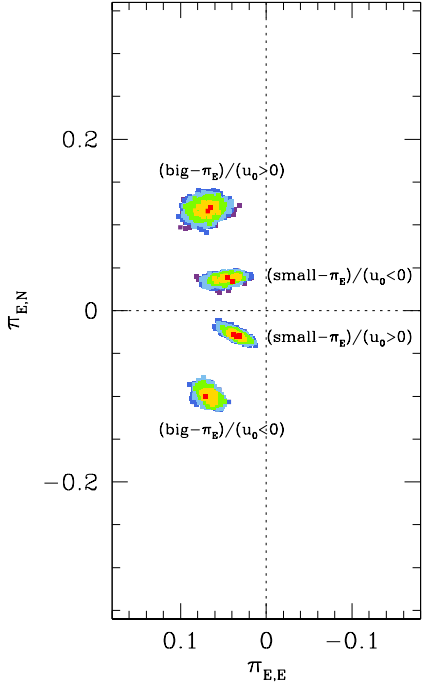}
\caption{
Distribution of $\Delta\chi^2$ in the ($\pi_{{\rm E},N},\pi_{{\rm E},E}$) plane obtained
from the analysis based on the ground+$Spitzer$ data. Color coding is same as in 
Fig.~\ref{fig:two}.  The local minima indicate the positions of the 4 degenerate solutions.
}
\label{fig:three}
\end{figure}

\begin{deluxetable*}{lcccc}
\tablecaption{Best-fit parameters (with $Spitzer$ data)\label{table:two}}
\tablewidth{0pt}
\tabletypesize{\small}
\tablehead{
\multicolumn{1}{c}{Parameter}     &
\multicolumn{2}{c}{Small $\pie$}  &
\multicolumn{2}{c}{Big $\pie$}     \\
\multicolumn{1}{c}{}     &
\multicolumn{1}{c}{$u_0<0$}     &
\multicolumn{1}{c}{$u_0>0$}     &
\multicolumn{1}{c}{$u_0<0$}     &
\multicolumn{1}{c}{$u_0>0$}     
}
\startdata                                              
$\chi^2$                    &  2373.1 (3.1)               &  2398.4 (12.7)            &   2395.1 (5.2)             &   2384.4 (1.7)            \\   
$t_0$ (HJD')                &  7904.908 $\pm$ 0.098       &  7904.885 $\pm$ 0.106     &   7904.873 $\pm$ 0.057     &   7905.071 $\pm$ 0.057    \\   
$u_0$                       &  -0.151 $\pm$ 0.002         &  0.152 $\pm$ 0.002        &   -0.151 $\pm$ 0.001       &   0.149 $\pm$ 0.001       \\   
$t_{\rm E}$ (days)          &  41.73 $\pm$ 0.06           &  41.73 $\pm$ 0.05         &   41.71 $\pm$ 0.04         &   41.64 $\pm$ 0.04        \\   
$s$                         &  1.438 $\pm$ 0.002          &  1.438 $\pm$ 0.002        &   1.440 $\pm$ 0.001        &   1.438 $\pm$ 0.001       \\   
$q$                         &  0.704 $\pm$ 0.005          &  0.702 $\pm$ 0.006        &   0.701 $\pm$ 0.002        &   0.712 $\pm$ 0.003       \\   
$\alpha$ (rad)              &  -0.642 $\pm$ 0.001         &  0.642 $\pm$ 0.001        &   -0.648 $\pm$ 0.001       &   0.647 $\pm$ 0.001       \\   
$\rho$  (10$^{-3}$)         &  8.76 $\pm$ 0.07            &  8.69 $\pm$ 0.07          &   8.68 $\pm$ 0.05          &   8.78 $\pm$ 0.06         \\   
$\pi_{{\rm E},N}$           &  0.034 $\pm$ 0.003          &  -0.030 $\pm$ 0.004       &   -0.100 $\pm$ 0.006       &   0.121 $\pm$ 0.007       \\   
$\pi_{{\rm E},E}$           &  0.040 $\pm$ 0.009          &  0.031 $\pm$ 0.007        &   0.070 $\pm$ 0.007        &   0.065 $\pm$ 0.009       \\   
$ds/dt$       (yr$^{-1}$)   &  0.215 $\pm$ 0.058          &  0.197 $\pm$ 0.059        &   0.122 $\pm$ 0.013        &   0.175 $\pm$ 0.021       \\   
$d\alpha/dt$  (yr$^{-1}$)   &  -0.165 $\pm$ 0.031         &  0.158 $\pm$ 0.030        &   -0.168 $\pm$ 0.020       &   0.009 $\pm$ 0.019       \\   
$f_{{\rm S},I}$             &  7.35                       &  7.36                     &   7.37                     &   7.35                    \\   
$f_{{\rm b},I}$             &  -0.06                      &  -0.07                    &   -0.08                    &   -0.06                  
\enddata                                              
\tablecomments{ 
The values in the parenthesis of the $\chi^2$ line represent the penalty $\chi^2$ values given by the color constraint.
See more details in section 3.2. ${\rm HJD}'={\rm HJD}-2450000$.\\
}
\end{deluxetable*}

\subsection{Additional Space-based Data}\label{sec:three-two}

Knowing the difficulty of secure $\pie$ measurement based on only the ground-based data, we test the 
possibility of $\pie$ measurement with the additional data obtained from $Spitzer$ observations. To compute 
lensing magnifications seen from the $Spitzer$ telescope, one needs the position and the distance to the 
satellite. The position of the $Spitzer$ telescope was in the ranges of 
$110^\circ \lesssim {\rm RA}\lesssim  194^\circ$ and $-7^\circ\lesssim {\rm DEC}\lesssim 21^\circ$ and the 
distance was in the range of $1.566\lesssim d_{\rm sat}/{\rm au}\lesssim 1.584$ during the 2017 bulge season.

The $Spitzer$ data partially covered the event light curve.  Furthermore, they do not cover major features 
such as those produced by caustic crossings. See the $Spitzer$ data presented in Figure~\ref{fig:one}.  In 
such a case, it is known that external information of the color between the passbands used for observations 
from Earth and from the $Spitzer$ telescope can be useful in finding a correct model \citep{Yee2015, Shin2017}. 
We, therefore, apply a color constraint with the measured 
instrumental color of $I-L=2.33\pm 0.012$
The color constraint 
is imposed by giving penalty $\chi^2$ defined in Eq. (2) of \citet{Shin2017}.

For single-lensing events observed both from Earth and from a satellite, it is known that there exist four 
sets of degenerate solutions \citep{Refsdal1966, Gould1994}. This degeneracy among these solutions, referred 
to as $(+,+)$, $(-,-)$, $(+,-)$, and $(-,+)$ solutions, arises due to the ambiguity in the signs of the 
lens-source impact parameters as seen from Earth (the former sign in the parenthesis) and from the satellite 
(the latter sign in the parenthesis).  In many case of binary-lens events, this four-fold degeneracy reduces 
into a two-fold degeneracy \citep{Han2017} due to the lack of lensing magnification symmetry. The remaining 
two degenerate solutions, $(+,+)$ and $(-,-)$ solutions, are caused by the mirror symmetry of the source 
trajectory with respect to the binary axis, and thus the degeneracy corresponds to the `ecliptic degeneracy'. 
Besides the known types, binary events can be subject to various other types of degeneracy.

In order to check the existence of degenerate solutions, we explore the space of the lensing parameters 
using two methods.  First, we conduct a grid search over the ($\pi_{{\rm E},N},\pi_{{\rm E},E}$) plane.  
Second, we search for local solutions using a downhill approach from various starting points that are 
obtained from the analysis based on the ground-based data. From these searches, we identify 2 classes of 
solutions, in which each class is composed of 2 solutions arising from the ecliptic degeneracy.

\begin{figure}
\includegraphics[width=\columnwidth]{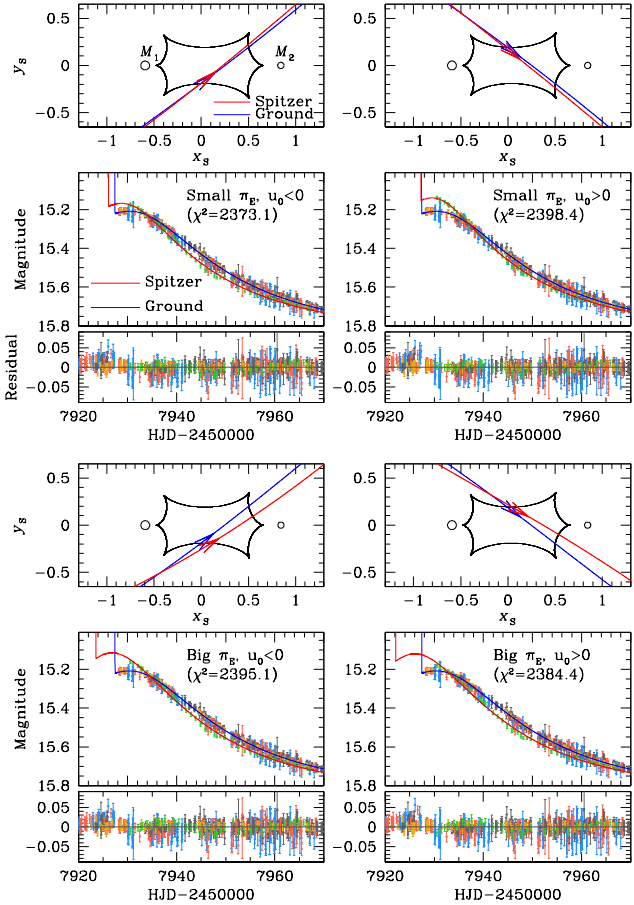}
\caption{
The lens system geometry and the portion of the light curve in the vicinity of the $Spitzer$ data for the 
4 degenerate 
solutions.  For each lens system geometry, the source trajectories seen from Earth and the $Spitzer$ telescope 
are marked by blue and red curves (with arrows), respectively. The cuspy closed curve represents the caustic.  
The coordinates are centered at the barycenter of the binary lens.  The blue and red curves superposed on the 
data points represent the model light curves for the ground and $Spitzer$ data, respectively.  
}
\label{fig:four}
\end{figure}

In Figure~\ref{fig:three}, we present the locations of the local solutions in the
($\pi_{{\rm E},N},\pi_{{\rm E},E}$) plane. It is found that one pair of solutions have 
$\pi_{\rm E}=(\pi_{{\rm E},N}^2 + \pi_{{\rm E},E})^{1/2}\gtrsim 0.1$ (`big-$\pie$' solutions) and the other 
pair have $\pie\lesssim 0.1$ (`small-$\pie$' solutions).  For each pair, the lensing parameters of the two 
degenerate solutions are approximately in the relation of 
$(u_0, \alpha, \pi_{{\rm E},N}, d\alpha/dt) \leftrightarrow -(u_0, \alpha, \pi_{{\rm E},N}, d\alpha/dt)$,
and thus we refer to the solutions as $u_0<0$ and $u_0>0$ solutions.  We note that although the higher-order 
parameters $(\pi_{{\rm E},N}, \pi_{{\rm E},E}, ds/dt, d\alpha/dt)$ of these degenerate solutions are different 
from one another, the other lensing parameters are similar because they are mostly determined from the 
ground-based data.  By comparing the ranges of the $\Delta\chi^2$ distributions with and without the 
{\it Spitzer} data, one finds that the uncertainties of the determined microlens-parallax parameters 
are greatly reduced with the use of the $Spitzer$ data.

In Table~\ref{table:two}, we list the lensing parameters of the 4 degenerate solutions along with their 
$\chi^2$ values. We find that the (small-$\pie$)/($u_0<0$) solution is preferred over the other solutions 
for two major reasons.  First, the (small-$\pie$)/($u_0<0$) solution provides a better fit to the observed 
data than the other solutions by $11.3 < \Delta\chi^2 <25.3$. Second, the small-$\pie$ solutions are 
preferred over the big-$\pie$ solution according to the ``Rich argument'' \citep{Calchi2015a}, which 
states that, other factors being equal, small parallax solutions are preferred over large ones by a 
probability factor $(\pi_{\rm E, big}/\pi_{\rm E,small})^2\gtrsim 6$. Although the (small-$\pie$)/($u_0<0$) 
solution is favored, one cannot completely rule out the other degenerate solutions. We, therefore, 
discuss the methods that can firmly resolve the degeneracy in Section~\ref{sec:five}.  Also presented in 
Table~\ref{table:two} are the fluxes of the source, $f_{{\rm S},I}$, and the blend, $f_{{\rm b},I}$, 
estimated based on the OGLE data.  
The small $f_{{\rm b},I}$ indicates that the blend flux is small.  
We note that the small negative blending is quite common for point-spread-function photometry 
in crowded fields \citep{Park2004}.

\begin{figure}
\includegraphics[width=\columnwidth]{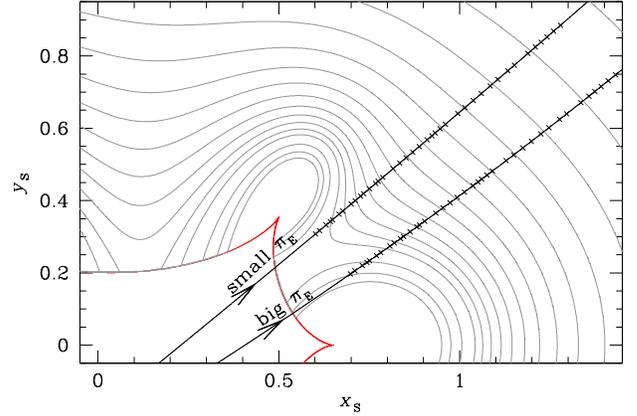}
\caption{
Contour map of lensing magnification in the outer region of the caustic. Contours are drawn at every 
$\Delta A=0.05$ step from $A=1.1$ to $A=2.0$. The lines with arrows represent the source trajectories 
of the small-$\pie$ and big-$\pie$ solutions seen from the $Spitzer$ telescope.  The crosses on each 
trajectory represent the expected positions of the source when $Spitzer$ data were taken.
}
\label{fig:five}
\end{figure}

In Figure~\ref{fig:four}, we present the lens-system geometry of the 4 degenerate solutions.  For each 
geometry, we present the source trajectories with respect to the lens components (small empty dots marked 
by $M_1$ and $M_2$) and the caustic (cuspy closed curve).  For each geometry, the source trajectories seen 
from Earth and the $Spitzer$ telescope are marked by blue and red curves (with arrows), respectively.  We 
also present the portion of the light curve in the vicinity of the $Spitzer$ data and the model light curve.

As mentioned, 
the degeneracy between the $u_0<0$ and $u_0>0$ solutions is caused by the mirror symmetry of the lens system 
geometry. On the other hand, the degeneracy between the small-$\pie$ and big-$\pie$ solutions is caused by the 
difference in the source trajectory angles seen from the $Spitzer$ telescope.  For the small-$\pie$ solution, 
the source trajectory angle as seen from the $Spitzer$ telescope is bigger than the angle of the source trajectory 
seen from the ground. In contrast, the $Spitzer$ trajectory angle of the big-$\pie$ solution is smaller than the 
angle of the ground trajectory.  We note that the latter degeneracy is different from the degeneracy between $(+,+)$ 
and $(+,-)$ solutions because both ground and satellite trajectories pass on the same side with respect to the 
barycenter of the binary lens. Such a degeneracy is not previously known.

In order to further investigate the cause of the degeneracy between the small-$\pie$ and big-$\pie$ solutions, 
in Figure~\ref{fig:five}, we present the magnification contours in the outer region of the caustic.  On the 
contour map, we plot the $Spitzer$ source trajectories of the two degenerate solutions.  From the map, it is 
found that the magnification patterns along the source trajectories of the two degenerate solutions are similar 
to each other, suggesting that the degeneracy is caused by the symmetry of magnification pattern in the outer 
region of the caustic.  We note that the degeneracy could have been resolved if the caustic exit part of the 
light curve had been covered by $Spitzer$ data because the times of the caustic exit (seen from the $Spitzer$ 
telescope) expected from the two degenerate solutions are different from each other.   
We find that the caustic-exit times for the small-$\pie$ solutions are in the range of  
$7926~({\rm for}~u_0<0) \lesssim  {\rm HJD}' \lesssim 7928~({\rm for}~u_0>0)$.
On the other hand, the range for the big-$\pie$ solutions is
$7922~({\rm for}~u_0<0) \lesssim  {\rm HJD}' \lesssim 7924~({\rm for}~u_0>0)$.
With the $\sim 4$ day time gap between the caustic-crossing times of the small-$\pie$ and big-$\pie$ 
solutions, the degeneracy could have been easily lifted. 
In conclusion, we find that the new type of degeneracy is caused by 
the partial symmetry of the magnification pattern outside the caustic combined with the fragmentary coverage 
of the $Spitzer$ data.

From the comparison of the analyses conducted with and without the space-based data, we find two
important results. 
\begin{enumerate}
\item
First, while the microlens parallax cannot be securely determined based on only the ground-based data, the 
addition of the $Spitzer$ data enables us to clearly identify two classes of microlens-parallax solutions. 
The degeneracy is either intrinsic to lensing systems ($u_0<0$ versus $u_0>0$ degeneracy) or due to the 
combination of the partial symmetry of magnification pattern combined with the fragmentary $Spitzer$ coverage 
of the event (small-$\pie$ versus big-$\pie$ degeneracy). 
\item
Second, the space-based data greatly improve the precision of the $\pie$ measurement. We find that the 
measurement uncertainties of the north and east components of $\pivec_{\rm E}$ are reduced by factors 
$\sim 18$ and $\sim 4$, respectively, with the use of the $Spitzer$ data. Since the lens mass is directly 
proportional to $1/\pie$, the uncertainty of the mass measurement reduces by the same factors.
\end{enumerate}

\begin{deluxetable*}{lcccc}
\tablecaption{Physical lens parameters\label{table:three}}
\tablewidth{0pt}
\tabletypesize{\normalsize}
\tablehead{
\multicolumn{1}{c}{Parameter}     &
\multicolumn{2}{c}{Small $\pie$}  &
\multicolumn{2}{c}{Big $\pie$}     \\
\multicolumn{1}{c}{}            &
\multicolumn{1}{c}{$u_0<0$}     &
\multicolumn{1}{c}{$u_0>0$}     &
\multicolumn{1}{c}{$u_0<0$}     &
\multicolumn{1}{c}{$u_0>0$}     
}
\startdata                                              
$M_1$ ($M_\odot$)            & 1.09 $\pm$ 0.15  &  1.33 $\pm$ 0.21       &  0.47 $\pm$ 0.04   &   0.41 $\pm$  0.04    \\   
$M_2$ ($M_\odot$)            & 0.77 $\pm$ 0.11  &  0.93 $\pm$ 0.15       &  0.33 $\pm$ 0.03   &   0.29 $\pm$  0.03    \\   
$D_{\rm L}$ (kpc)            & 6.38 $\pm$ 0.79  &  6.66 $\pm$ 0.86       &  4.69 $\pm$ 0.45   &   4.48 $\pm$  0.42    \\   
$a_\perp$ (au)               & 7.20 $\pm$ 0.89  &  7.59 $\pm$ 0.98       &  5.35 $\pm$ 0.52   &   5.04 $\pm$  0.47    \\   
$({\rm KE}/{\rm PE})_\perp$  & 0.12 $\pm$ 0.02  &  0.12 $\pm$ 0.02       &  0.09 $\pm$ 0.01   &   0.04 $\pm$  0.01    \\
$\psi$ (deg)                 & $51$             &  $132$                 &  $141$             &  $32$
\enddata                                                
\end{deluxetable*}

\section{Physical Lens Parameters}\label{sec:four}

\subsection{Source Star and Angular Einstein Radius}\label{sec:four-one}

For the unique determination of the lens mass and distance, one needs to estimate the angular Einstein 
radius in addition to the microlens parallax.  Since the angular Einstein radius is determined by 
$\thetae=\theta_*/\rho$, it is required to estimate the angular radius of the source star.

\begin{figure}
\includegraphics[width=\columnwidth]{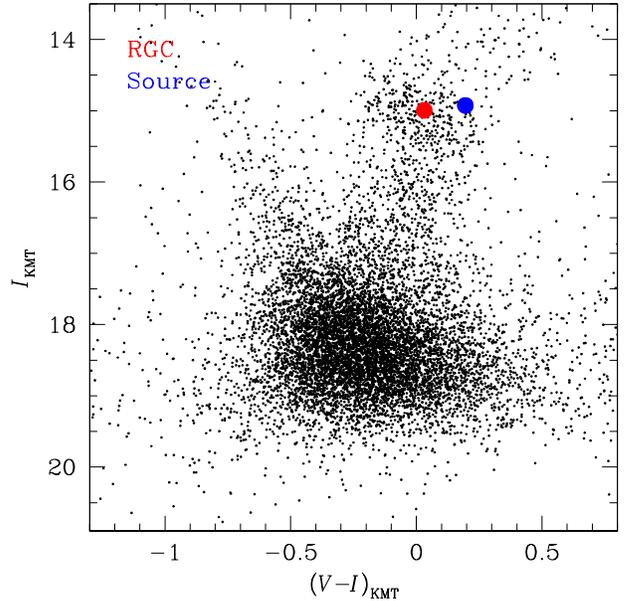}
\caption{
Location of the source with respect to the centroid of red giant clump (RGC) in the instrumental 
color-magnitude diagram.
}
\label{fig:six}
\end{figure}

We estimate $\theta_*$ from the dereddened color $(V-I)_0$ and brightness $I_0$ of the source. In order to 
derive $(V-I)_0$ from the instrumental color-magnitude diagram, we use the method of \citet{Yoo2004}, which 
uses the centroid of red giant clump (RGC) as a reference. In Figure~\ref{fig:six}, we present the location 
of the source and the RGC centroid in the instrumental color-magnitude diagram constructed from the $I$- and $V$-band DoPHOT 
photometry of the KMTC data set. It is found that the offsets in color and brightness of the source with respect 
to the RGC centroid are $\Delta (V-I,I)=(0.16,-0.07)$. With the known dereddened color and magnitude of RGC centroid, 
$(V-I,I)_{\rm RGC}=(1.06,14.5)$ \citep{Bensby2011, Nataf2013}, we find that the dereddened color and brightness of 
the source star are $(V-I,I)_0= (V-I,I)_{\rm RGC}+\Delta(V-I,I)=(1.22\pm 0.07,14.48\pm 0.09)$.  This indicates that 
the source is a K-type giant star. Using the color-color relation provided by \citet{Bessell1988}, we convert $V-I$ 
into $V-K$. Using the relation between $V-K$ and the surface brightness \citep{Kervella2004}, we estimate the angular 
source radius. The estimated angular source radius is $\theta_* = 6.9 \pm 0.6\ \mu{\rm as}$.  Combined with value of 
$\rho$, we estimate that  the angular Einstein radius of the lens system is
\begin{equation}
\thetae = 0.79 \pm 0.06\ {\rm mas}. 
\end{equation}
With the measured angular Einstein radius, the relative lens-source proper motion is estimated by
\begin{equation}
\mu = 6.89 \pm 0.56\ {\rm mas}\ {\rm yr}^{-1}.
\end{equation}

\subsection{Lens Parameters}\label{sec:four-two}

With the measured microlens parallax and the angular Einstein radius, we estimate the mass and distance to the 
lens using the relations in Equation~(2). In Table~\ref{table:three}, we list the determined physical parameters, 
including masses of the primary, $M_1$, and the companion, $M_2$, the distance to the lens, $D_{\rm L}$, and the 
projected separation between the lens components, $a_\perp=sD_{\rm L}\thetae$, for the individual degenerate 
lensing solutions.  To check the physical validity of the parameters, we also present the ratio between the 
projected potential energy to the kinetic energy, i.e., $({\rm KE}/{\rm PE})_\perp$.

Due to the difference in the microlens-parallax values between the small-$\pie$ and big-$\pie$ solution classes, 
the estimated lens masses and distances for the two classes of solutions are substantially different from each 
other.  On the other hand, the physical parameters for the pair of the $u_0>0$ and $u_0<0$ solutions are similar 
to each other.  We find that the masses of the primary and companion are $1.1\lesssim M_1/M_\odot \lesssim 1.3$ 
and $0.8\lesssim M_2/M_\odot \lesssim 0.9$ for the small-$\pie$ solutions.  For the big-$\pie$ solutions, the 
masses of the lens components are $0.4\lesssim M_1/M_\odot \lesssim 0.5$ and $M_2\sim 0.3\ M_\odot$.  The 
estimated distances to the lens are $7.2\lesssim D_{\rm L}/{\rm kpc}\lesssim 7.6$ and 
$5.0\lesssim D_{\rm L}/{\rm kpc}\lesssim 5.4$ according to the small-$\pie$ and big-$\pie$ solutions, respectively.

\begin{deluxetable}{lccccc}
\tablecaption{Expected Lens Brightness\label{table:four}}
\tablewidth{0pt}
\tabletypesize{\small}
\tablehead{
\multicolumn{2}{c}{Solution}     &
\multicolumn{2}{c}{Lens}  &
\multicolumn{2}{c}{Source}     \\
\multicolumn{1}{c}{}            &
\multicolumn{1}{c}{}            &
\multicolumn{1}{c}{$I_{\rm L}$}     &
\multicolumn{1}{c}{$H_{\rm L}$}     &
\multicolumn{1}{c}{$I_{\rm S}$}     &
\multicolumn{1}{c}{$H_{\rm S}$}     
}
\startdata                                              
Small $\pie$  &  $u_0<0$  & 18.8  &   17.0     & 15.8   &  13.5  \\   
              &  $u_0>0$  & 17.7  &   16.5     &  -     &  -     \\   
Big $\pie$    &  $u_0<0$ & 21.5   &   18.7     &  -     &  -     \\   
              &  $u_0>0$ & 22.1   &   19.1     &  -     &  -          
\enddata                                              
\color{black}
\end{deluxetable}

\section{Resolving Degeneracy}\label{sec:five}

\subsection{Lens Brightness}

The estimated masses of the lens for the small-$\pie$ and big-$\pie$ solutions are considerably different due to 
the difference in the measured microlens-parallax values. Then, if the lens-source can be resolved from future
high-resolution imaging observations, the degeneracy can be resolved from the lens brightness.

If the proposed follow-up high-resolution observations are conducted, the observations will likely be conducted
in the near-IR band. We, therefore, estimate the $H$-band magnitudes of the source and lens. From the dereddened 
$I$-band magnitude $I_0\sim 14.5$, the dereddened $H$-band source magnitude of the source is $H_0\sim 13.1$ 
\citep{Bessell1988}.
The $V$-band extinction of $A_V\sim 2.6$ in combination with the extinction ratio $(A_H/A_V)\sim 0.108$ 
\citep{Nishiyama2008} toward the bulge field yields the $H$-band extinction of $A_H\sim 0.28$. Then, the apparent 
$H$-band magnitude of the source is $H_{\rm S}=H_0 + A_H\sim 13.4$. We compute the lens brightness based on the 
mass and distance under the assumption that the lens and source experience the same amount of extinction. In 
Table~\ref{table:four}, we present the expected combined (primary plus companion) $I$- and $H$-band magnitudes 
of the lens and source. The brightness of the lens varies depending on the solution. For the small-$\pie$ 
solutions, the apparent $H$-band magnitude of the lens is $H_{\rm L}\sim 16.5$ -- 17.0. For the big-$\pie$ 
solutions, on the other hand, the expected $H$-band lens brightness is $H_{\rm L}\sim 18.7$ --19.1.

According to the estimated $I$-band lens brightness, the lens-to-source flux ratio for the (small-$\pie$)/($u_0>0$) 
solution is $f_{{\rm L},I}/f_{{\rm S},I}\sim 17\%$.  Since the light from the lens contributes to blended light, 
then, this ratio is too big to be consistent with the small amount of the measured blended flux, even considering 
the uncertainties of the lens mass and distance.  Therefore, the solution is unlikely to be the correct solution 
not only because of its worst $\chi^2$ value among the degenerate solutions but also because of the limits on 
blended light.
The lens-to-source flux ratio for the (small-$\pie$)/($u_0<0$) is about 6\%,  
but with the lens mass and distance at the $1\sigma$  ($2\sigma$) level,
the ratio is $\sim 2\%$ ($1\%$), which is consistent with the blending.

\subsection{Relative Lens-source Proper Motion}

The degeneracy between $u_0<0$ and $u_0>0$ solutions can also be lifted once the lens and source are resolved.
The relative lens-source proper motion vector is related to 
$\te$, $\thetae$, and $(\pi_{{\rm E},N}, \pi_{{\rm E},E})$ by
\begin{equation}
\muvec = (\mu_N,\mu_E)=
\left( 
{\thetae\over t_{\rm E}} {\pi_{{\rm E},N} \over  \pie},
{\thetae\over t_{\rm E}} {\pi_{{\rm E},E} \over  \pie}
\right). 
\end{equation}
For the pair of the degenerate solutions with $u_0<0$ and $u_0>0$, the north components of $\pivec_{\rm E}$ have 
opposite signs. This implies that the relative motion vectors of the two degenerate solutions are directed in 
substantially different directions
and thus the degeneracy can be resolved from the lens motion with respect to the source.

The heliocentric lens-source proper motion is $\mu_{\rm helio}\sim  7~{\rm mas}~{\rm yr}^{-1}$, 
which is about what is expected for a disk lens.
In this case, the expected direction of $\muvec_{\rm helio}$ (i.e., the direction of
Galactic rotation $\psi\sim 30^\circ$) is roughly $30^\circ$ East of North.  
In Table~\ref{table:three}, we list the orientation angles $\psi$ of $\muvec_{\rm helio}$, 
as measured from North to East, corresponding to the individual solutions.  
The heliocentric proper motion is related to the geocentric proper motion $\muvec_{\rm geo}$ by
\begin{equation}
\muvec_{\rm helio}=\muvec_{\rm geo}+{\bf v}_{\oplus,\perp}{ \pi_{\rm rel}\over {\rm au}},
\end{equation}
where ${\bf v}_{\oplus,\perp}$ represents the projected Earth motion at $t_0$.  One finds that the expected 
direction $\muvec_{\rm helio}$ is consistent with the (small-$\pie$)/($u_0<0$) and the (big-$\pie$)/($u_0>0$) 
solutions but inconsistent with the others.

From Keck adaptive optics observations, \citet{Batista2015} resolved the lens from the source $\sim 8.2$ years
after the event OGLE-2005-BLG-169, for which the relative lens-source proper motion is 
$\mu \sim 7.4\ {\rm mas}\ {\rm yr}^{-1}$.
The estimated proper motion of OGLE-2017-BLG-0329 ($\mu \sim 6.9\ {\rm mas}\ {\rm yr}^{-1}$) is similar to that 
of OGLE-2005-BLG-169. Considering the large lens/source flux ratio, the lens-source resolution by Keck
observations will take $\sim 10$ years, which is somewhat longer than the time for OGLE-2005-BLG-169. We
note that GMT/TMT/ELT, which will have better resolution than Keck, may be available before Keck can
resolve the event and thus the time for follow-up observations can be shortened.

\section{Conclusion}\label{sec:six}

We presented the analysis of the binary microlensing event OGLE-2017-BLG-0329, which was observed both 
from the ground and in space using the $Spitzer$ telescope. We found that the parallax model based on the 
ground-based data could not be distinguished from a zero-$\pie$ model at 2$\sigma$ level. However, with 
the use of the additional $Spitzer$ data, we could identify 2 classes of microlens-parallax solutions, 
each composed of a pair of solutions according to the well-known ecliptic degeneracy. We also found that 
the space-based data helped to greatly reduce the measurement uncertainties of the microlens-parallax 
vector $\pivec_{\rm E}$. With the measured microlens parallax combined with the angular Einstein radius 
measured from the resolved caustics, we found that the lens was composed of a binary with components 
masses of either $(M_1,M_2)\sim (1.1,0.8)\ M_\odot$ or $\sim (0.4,0.3)\ M_\odot$ according to the two 
solution classes.  The degeneracy among the solution would be resolved from adaptive optics observations 
taken $\sim 10$ years after the event.

\acknowledgments
Work by C.~Han was supported by the grant (2017R1A4A1015178) of
National Research Foundation of Korea.
Work by WZ, YKJ, and AG were supported by AST-1516842 from the US NSF.
WZ, IGS, and AG were supported by JPL grant 1500811.
The OGLE project has received funding from the National Science Centre, Poland, grant 
MAESTRO 2014/14/A/ST9/00121 to A.~Udalski.  
Work by YS was supported by an appointment to the NASA Postdoctoral Program at the Jet
Propulsion Laboratory, administered by Universities Space Research Association
through a contract with NASA.
This work was (partially) supported by NASA contract NNG16PJ32C.
Work by S.~Rahvar and S.~Sajadian is supported by INSF-95843339.
This research has made use of the KMTNet system operated by the Korea
Astronomy and Space Science Institute (KASI) and the data were obtained at
three host sites of CTIO in Chile, SAAO in South Africa, and SSO in
Australia.
We acknowledge the high-speed internet service (KREONET)
provided by Korea Institute of Science and Technology Information (KISTI).


\begin{thebibliography}{}

\bibitem[Alard \& Lupton(1998)]{Alard1998} Alard, C., \& Lupton, R.~H.\ 1998, \apj, 503, 325
\bibitem[Batista et al.(2015)]{Batista2015} Batista, V., Beaulieu, J.-P., Bennett, D.~P., et al.\ 2015, \apj, 808, 170
\bibitem[Batista et al.(2011)]{Batista2011} Batista, V., Gould, A., Dieters, S., et al.\ 2011, \aap, 529, 102
\bibitem[Bensby et al.(2011)]{Bensby2011} Bensby, T., Ad\'en, D., Mel\'endez, J., et al.\ 2011, \pasp, 533, 134
\bibitem[Bessell \& Brett(1988)]{Bessell1988} Bessell, M.~S., \& Brett, J.~M.\ 1988, \pasp, 100, 1134
\bibitem[Calchi Novati et al.(2015a)]{Calchi2015a} Calchi Novati, S., Gould, A., Udalski, A., et al.\ 2015a, \apj, 804, 20
\bibitem[Calchi Novati et al.(2015b)]{Calchi2015b} Calchi Novati, S., Gould, A., Yee, J.~C., et al.\ 2015b, \apj, 814, 92
\bibitem[Gould(1992)]{Gould1992} Gould, A.\ 1992, \apj, 392, 442
\bibitem[Gould(1994)]{Gould1994} Gould, A. 1994, \apjl, 421, L71
\bibitem[Gould(1994)]{Gould1994} Gould, A.\ 1994, \apjl, 421, L75
\bibitem[Gould(2000)]{Gould2000} Gould, A.\ 2000, \apj, 542, 785
\bibitem[Han \& Gould(1995)]{Han1995} Han, C., \& Gould, A. 1995, \apj, 447, 53
\bibitem[Han et al.(2016)]{Han2016} Han, C., Udalski, A., Gould, A., et al.\ 2016, \aj, 152, 95
\bibitem[Han et al.(2017)]{Han2017} Han, C., Udalski, A., Gould, A., et al.\ 2017, \apj, 834, 82
\bibitem[Kayser et al.(1986)]{Kayser1986} Kayser, R., Refsdal, S., \& Stabell, R.\ 1986, \aap, 166, 36
\bibitem[Kervella et al.(2004)]{Kervella2004} Kervella, P., Th\'evenin, F., Di Folco, E., \& S\'egransan, D.\ 2004, \aap, 426, 297
\bibitem[Kim et al.(2016)]{Kim2016} Kim, S.-L., Lee, C.-U., Park, B.-G., et al.\ 2016, JKAS, 49, 37
\bibitem[Nataf et al.(2013)]{Nataf2013} Nataf, D.~M., Gould, A., Fouqu\'e, P., et al.\ 2013, \apj, 769, 88
\bibitem[Nemiroff \& Wickramasinghe(1994)]{Nemiroff1994} Nemiroff, R.~J., \& Wickramasinghe, W.~A.~D.~T. 1994, \apj, 424, L21
\bibitem[Nishiyama et a.(2008)]{Nishiyama2008} Nishiyama, S., Nagata, T., Tamura, M., Kandori, R., Hatano, H., Sato, S., \& Sugitani, K.\ 2008, \apj, 680, 1174
\bibitem[Park et al.(2004)]{Park2004} Park, B.-G., DePoy, D.~L., Gaudi, B.~S., et al.\ 2004, \apj, 609, 166
\bibitem[Refsdal(1966)]{Refsdal1966} Refsdal, S.\ 1966, \mnras, 134, 315
\bibitem[Schneider \& Weiss(1986)]{Schneider1986} Schneider, P., \& Weiss, A.\ 1986, \aap, 164, 237
\bibitem[Shin et al.(2017)]{Shin2017} Shin, I.-G., Udalski, A., Yee, J. C., et al.\ 2017, \aj, 154, 176
\bibitem[Skowron et al.(2011)]{Skowron2011} Skowron, J., Udalski, A., Gould, A., et al.\ 2011, \apj, 738, 87
\bibitem[Smith et al.(2003)]{Smith2003} Smith, M.~C., Mao, S., \& Paczy\'nski, B,\ 2003, \mnras, 339, 925
\bibitem[Udalski et al.(2015)]{Udalski2015a} Udalski, A., Szyma\'nski, M.~K., \& Szyma\'mski, G.\ 2015, Acta Astron., 65, 1
\bibitem[Udalski et al.(2015)]{Udalski2015b} Udalski, A., Yee, J.~C., Gould, A., et al.\ 2015, \apj, 799, 237
\bibitem[Wambsganss(1997)]{Wambsganss1997} Wambsganss, J.\ 1997, \mnras, 284, 172
\bibitem[Wang et al.(2017)]{Wang2017} Wang, T., Zhu, W., Mao, S., et al.\ 2017, \apj, 845, 129
\bibitem[Witt \& Mao(1994)]{Witt1994} Witt, H. J., \& Mao, S. 1994, \apj, 430, 505
\bibitem[Yee et al.(2015)]{Yee2015} Yee, J.~C., Udalski, A., Calchi Novati, S., et al.\ 2015, \apj, 802, 76
\bibitem[Yee et al.(2012)]{Yee2012} Yee, J.~C., Shvartzvald, Y., Gal-Yam, A., et al. 2012, \apj, 755, 102 
\bibitem[Yoo et al.(2004)]{Yoo2004} Yoo, J., DePoy, D.~L., Gal-Yam, A., et al.\ 2004, \apj, 603, 139

\end{thebibliography}
\end{document}